\documentclass{article}
\usepackage[utf8]{inputenc}
\usepackage{amsmath}
\usepackage{lineno}
\usepackage{graphicx}
\usepackage{geometry}
\usepackage{graphicx}
\usepackage{caption}
\usepackage{subcaption}
\usepackage{comment}
\usepackage{natbib}
\usepackage{setspace}

\title{A Note on Planet Size and Cooling Rate}
\date{}
\author{Johnny Seales and Adrian Lenardic}
\begin{document}

\maketitle

\begin{abstract}
Variation in the balance of forces that drive and resist tectonic plate motions allows small terrestrial planet to cooler slower than larger ones. Given that interior cooling affects surface environment, through volcanic/geologic activity, this indicates that small planets should not be down-weighted in the search for life beyond Earth.
\end{abstract}

\emph{Keywords: planetary cooling, habitability, exoplanets }

\section{Introduction}
The idea that small planets cool faster than larger ones stems from an area to volume argument. For a planet of radius, $R_p$, heat flow scales with surface area while heat produced within its interior scales with volume. Taking the ratio, cooling scales as $1/R_p$. For scaling between planets, this assumes that planets have the same internal heat source concentrations, valid for planets of similar chemical composition. It also assumes equivalent surface heat flux for a given internal temperature. 

It has been noted that the relationship between heat flux and internal temperature depends on the tectonic mode of a planet. Planets with plate tectonics will have a different cooling efficiency than single plate planets \citep{Stevenson2003}. Although that potential has been acknowledged, there is still the thought that surface to volume arguments remain valid for planets with the same tectonic modes - in particular, plate tectonics. This assumption has not been called out to date, and it too is invalid. 

Plate tectonics is a kinematic theory \citep{McKenzie1967,Morgan1991,LePichon1968}. Connecting plate tectonics to interior cooling is a dynamic problem and a dynamic theory of plate tectonics is not agreed upon at present. It is an active research problem with no deficiency of hypotheses. Different assumptions regarding the balance between the forces driving and resisting plate motions lead to different scaling relationships between heat flux and internal temperature \citep{Tozer1972b,Christensen1985,Conrad1999b,Crowley2012}. Such relationships have been used in models that track the evolution of the Earth's internal temperature over time (thermal history models). 

Proponents of different hypotheses regarding the cooling efficiency of plate tectonics have argued that their models can match thermal history constraints. However, using an agnostic approach that accounted for model and observational uncertainties, \citet{Seales2019,Seales2020} showed that multiple hypotheses remain viable. More critically for planetary studies, no observational data demands equivalency between the cooling efficiency of planets within a tectonic mode akin to Earth and the Earth itself. In this note we explore how variances in plate tectonic cooling efficiencies couple with variable planetary size to determine cooling rates.

\section{Thermal History Models}
\indent  Plate tectonics, on Earth and potentially other terrestrial planets, is a surface manifestation of thermal convection within a planet's rocky interior layer (i.e., its mantle). Thermal history models predict mantle cooling trajectories based on how internal heat sources ($H$) and convective heat flux ($q_{conv}$) evolve with time. A large class of such models use a global energy balance that determines the spherically averaged temperature of the mantle, $T$, according to
\begin{equation} \label{Tdot}
\rho c_pV\dot{T}=H-Aq_{conv}
\end{equation}
where $rho$, $c_p$, and $\dot{T}$ are the density, heat capacitance and the time derivative of mantle temperature. The volume of the convecting mantle is $V=\frac{4\pi}{3}\left(R_p^3-R_c^3\right)$ and its surface area is $A=4\pi R_p^2$, where $R_c$ is the radius of the iron core of a terrestrial planet/moon. Radiogenic decay produces heat within the mantle according to 
\begin{equation} \label{H}
H=VH_oexp(-\lambda t)
\end{equation}
where $H_o$ is a scaling constant representing the heat produced per unit time per unit volume, $\lambda$ is the decay constant, and $t$ is time. The heat flux through the surface depends on convective vigor in the mantle. It is typically parameterized using a scaling equation given by \citet{Schubert1979,Schubert1980}:
\begin{equation}\label{eqparam}
Nu=aRa^{\beta}
\end{equation}
where $Nu$ is the Nusselt number (a measure of surface heat flux), $Ra$ is the Rayleigh number (a measure of convective vigor), $a$ is scaling constant that accounts for geometric effects (e.g., the wavelength of convection), and $\beta$ is a scaling exponent that encapsulates the efficiency of convective cooling. The value of $\beta$ varies between different hypotheses for the dynamics of plate tectonics. We will return to this issue after we develop the final model equations. The Nusselt number is the convective heat flux, $q_{conv}$, normalized by the amount of heat that would be conducted through the layer of thickness $D$. The conductive flux is given by Fourier's Law, $q_{cond}=\frac{k\Delta T}{D}$. The values $k$ and $\Delta T$ are the thermal conductivity and the difference between surface and interior temperatures. $Ra$ is  defined as
\begin{equation} \label{Ra}
Ra=\frac{\rho g\alpha\Delta T D^3}{\kappa \eta (T)}
\end{equation}
where $\rho$, $g$, $\alpha$ and $\kappa$ are density, gravity, thermal expansivity and thermal diffusivity. The temperature-dependent mantle viscosity is defined as
\begin{equation}
    \eta(T)=\eta_{ref}exp\left[\frac{A_e}{R}\left(\frac{1}{T}-\frac{1}{T_{ref}}\right)\right]
\end{equation}
where $A_e$ is the activation energy and $R$ the universal gas constant, and $eta_{ref}$ and $T_{ref}$ are reference values \citep{Karato1993}. Conduction becomes unstable when $Ra$ exceeds a critical threshold value, $Ra_c$. Taking this into account, the convective heat flux is
\begin{equation}\label{q_conv}
    q_{conv}=ak\frac{\Delta T}{D}\left(\frac{Ra}{Ra_c}\right)^\beta
\end{equation}
Combining the above we arrive at
\begin{equation}\label{CTM}
    \dot{T}=\frac{1}{\rho c_p}\left[H_oexp(-\lambda t) - \frac{A}{V}\frac{ak\Delta T}{D}\left(\frac{Ra}{Ra_c}\right)^\beta\right].
\end{equation}

If we assume that all values in Equation (\ref{Ra}) are constant except $T$ and $\eta (T)$, then combining Equations (\ref{eqparam}), (\ref{Ra}), and the definition of $Nu$ leads to
\begin{equation} \label{Q}
q_{conv}=a'\frac{T^{1+\beta}}{\eta (T)^{\beta}}
\end{equation}
\begin{equation} \label{a'}
a'=\frac{ak}{D}\left(\frac{\rho g \alpha D^3}{\kappa}\right)^\beta
\end{equation}
where all constants have now been combined into $a'$. The material constants can be determined using experimental values. The geometric constant, $a$, can be determined from laboratory and/or numerical convection experiments in combination with boundary layer theory \citep{Davies1980,Schubert1980}. We will refer to that approach as a classic thermal history model (CTM). An alternative approach, that we refer to as a scaled thermal history model (STM), sets the constant $a'$ to a particular heat flow, $q_o$, at a scaling temperature, $T_o$, and viscosity, $\eta_o$ \citep{Christensen1985}. In doing so, we have an alternative formulation given by 
\begin{equation}
    \dot{T}=\frac{1}{\rho c_p}\left[H_oexp(-\lambda t)-\frac{Aq_o}{V}\left(\frac{T}{T_o}\right)^{1+\beta}\left(\frac{\eta_o}{\eta (T)}\right)^\beta\right].
\end{equation}

CTMs integrate forwards in time from an initial mantle temperature value. STMs have historically built in Earth's present day heat flux, temperature, and viscosity directly into the model formulation (akin to a data assimilation approach). Following this rationale, STMs have integrated backwards in time to model past mantle temperatures starting from present day values. That is not conducive to modeling exoplanets, but the STM approach can be adapted for integrating forwards in time  \citep{Seales2019,Seales2020}. For completeness, we evaluated how variable planetary mass/size and tectonic cooling efficiency (i.e., different $\beta$ values) affected thermal histories by evolving model paths of both CTMs and STMs forwards in time.

\begin{table}[h]
\caption{Model constants, scaling values and parameter ranges}
\centering
\begin{tabular}{ c c c c }
\hline\hline
Symbol & Parameter & Value & Units    \\
\hline
$H_o$ & Initial radiogenic concentration & 1.25e-7 & $Wm^{-3}$ \\
$\lambda$ & Decay constant & 0.34  & $Gyr^{-1}$ \\
$\alpha$ & Thermal expansivity & 2e-5 & $K^{-1}$ \\
$\kappa$ & Thermal diffusivity & 1e-6 & $m^2s^{-1}$ \\
$T_s$ & Surface Temperature  & 273 & $K$ \\
$\eta_{ref}$ & Reference viscosity & $1e21 $ & $Pa*s$ \\
$A_e$ & Activation energy & 3e5 & $Jmol^{-1}$ \\
$R$ & Universal gas constant & 8.314 & $J(K*mol)^{-1}$ \\
$T_{ref}$ & Reference temperature & 1855 & $K$ \\
$Ra_c$ & Critical Rayleigh number & 1100 & - \\
$c_p$ & Heat capacitance & 1400 & $J(kg*K)^{-1}$ \\
$k$ & Thermal conductivity & 4.2 & $W(m\times K)^{-1}$ \\
$T_o$ & Scaling temperature & 1600 & $K$ \\
$q_o$ & Scaling convective heat flow & 0.069 & $Wm^{-2}$ \\
$\eta_o$ & Scaling viscosity & 4.45e19 & $Pa*s$ \\
$M_\oplus$ & Mass of Earth & 5.97e24 & $kg$ \\
$R_\oplus$ & Radius of Earth & 6371 & $km$ \\
$G$ & Gravitational constant & 6.67408e-11 & $Nm^2kg^{-2}$ \\
$\beta$ & Tectonic cooling efficiency constant & 0-0.33 & - \\
$M_p$ & Mass of Planet & 0.1-5 $M_\oplus$ & - \\
$R_p$ & Planet radius & Calculated & $km$ \\
$R_c$ & Core radius & Calculated & $km$ \\
$\rho$ & Mantle density & Calculated  & $kgm^{-3}$ \\
$g$ & Surface gravity & Calculated & $ms^{-2}$ \\
\hline
\end{tabular}
\label{table:convectiion_parameters}
\end{table}

The cooling efficiency of plate tectonics remains a matter of debate. For this reason, thermal history models have assumed different values of $\beta$. Given that different $\beta$ values represent different physical assumptions regarding the dynamics of plate tectonics, and by association Earth cooling, it follows that different values of $\beta$ represent different hypotheses. The earliest thermal history models used a $\beta$ value of 0.33 \citep{Schubert1980,Spohn1982,Jackson1984}. This assumes that mantle viscosity dominantly resists convective motion \citep{Tozer1972b}. \citet{Gurnis1989} incorporated analogues to tectonic plates and showed this scaling could be recovered provided that weak plate boundaries were also incorporated. \citet{Moresi1998} allowed weak plate boundaries to develop dynamically, which lead to a scaling exponent of 0.30. If plate boundaries are not assumed to be so weak that energy dissipation along them can be neglected and/or if plate strength offers resistance to convective motion, then the scaling exponent will be lower, with a range between $0<=\beta<=0.15$ having been proposed \citep{Christensen1985,Giannandrea1993,Conrad1999a,Conrad1999b}. \citet{Hoink2011} and \citet{Crowley2012} argued that different sized plates can have different balances between plate driving and resisting forces. This leads to a mixed mode scaling that allows for $\beta$ values between 0.15 and 0.30 \citep{Hoink2013}. We will consider the full range of $\beta$ values cited above. As noted in the introduction, within data and model uncertainties, multiple models within that range can match observational constraints on the cooling of the Earths interior over time \citep{Seales2019, Seales2020}.

Our choice of constants, scaling values and parameter ranges are listed in Table \ref{table:convectiion_parameters}. We calculated thermal paths for planets ranging from 0.1 to 5 earth masses ($M_\oplus$). For the remainder of this paper $\oplus$ refers to Earth-referenced values. For scaling models with planetary mass ($M_p$), we followed in the spirit of \citet{Schaefer2015} in using the scalings of \citet{Valencia2006} to determine the planetary ($R_p$) and core ($R_c$) radii, assuming a constant core mass fraction of 0.3259. We calculated the average mantle density ($\rho$) based on the planetary mass, mantle volume and the average gravitational acceleration ($g$), which scales as $GM_p/R_p^2$. We ran model suites with two different sets of initial temperatures. In one scenario, all planets began with the same average mantle temperature. The second suite of models started all planets with the same potential temperature - the temperature of the interior mantle removing the effects of adiabatic self-compression. We used the scaling of \citet{Schaefer2015} to convert between average mantle temperature and potential temperature. 

\begin{figure}[h!]
  \centering
    \begin{subfigure}[b]{0.35\textwidth}
      \includegraphics[width=\linewidth]{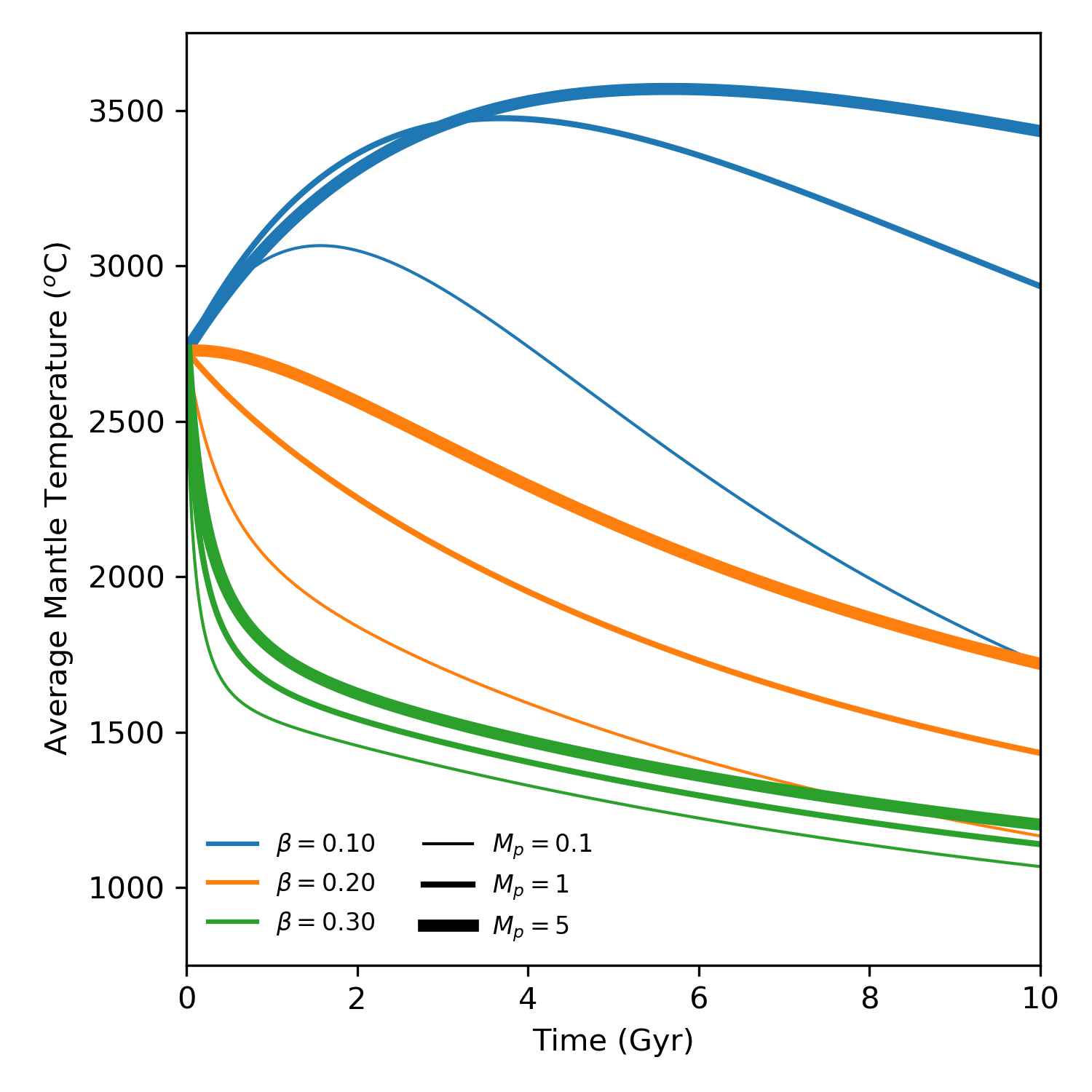}
      \caption{}
      \label{fig:Paths_SameTmi_classic}
    \end{subfigure}%
    \begin{subfigure}[b]{0.35\textwidth}
      \includegraphics[width=\linewidth]{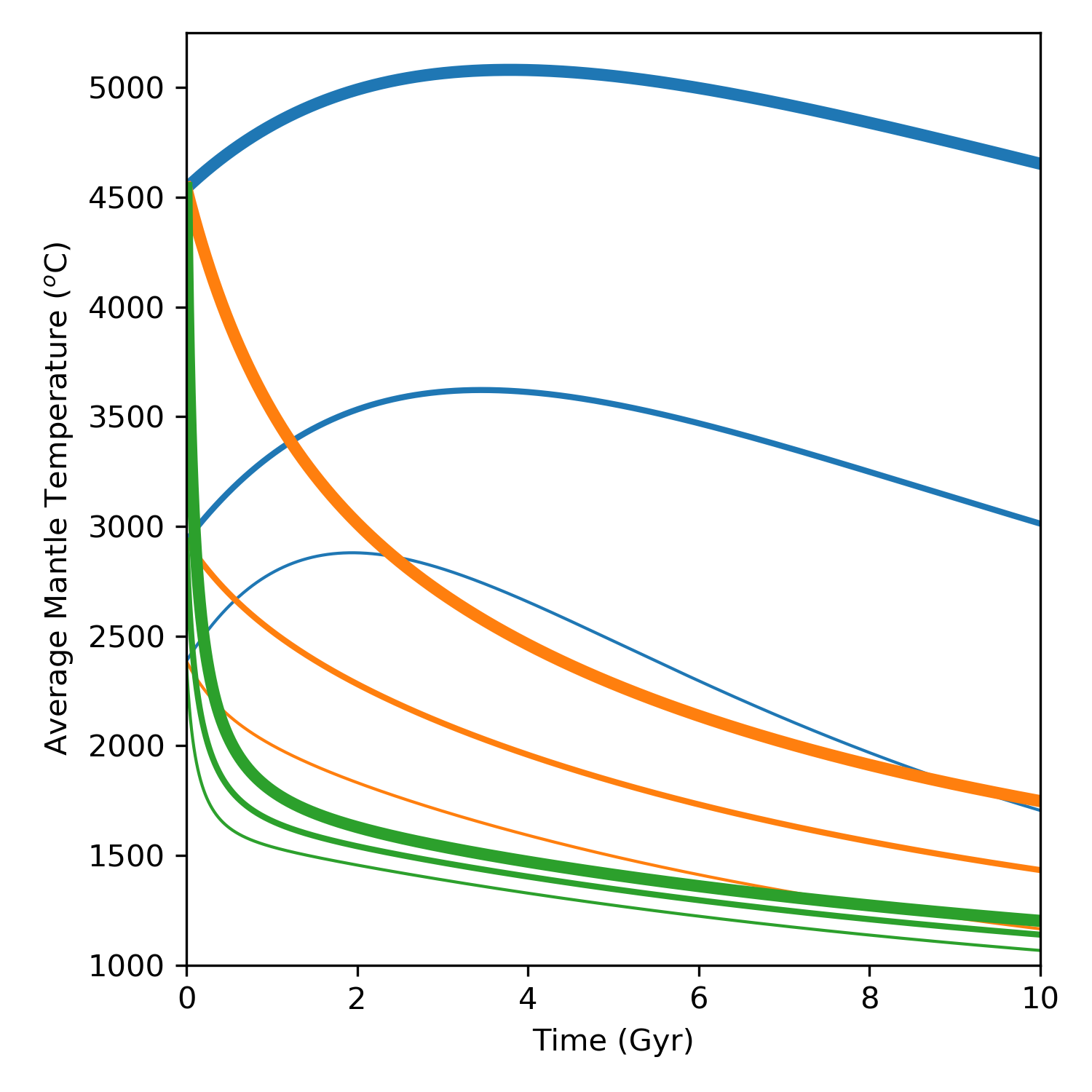}
      \caption{}
      \label{fig:Paths_DiffTmi_classic}
    \end{subfigure}%
    
    \medskip
    \begin{subfigure}[b]{0.35\textwidth}
      \includegraphics[width=\linewidth]{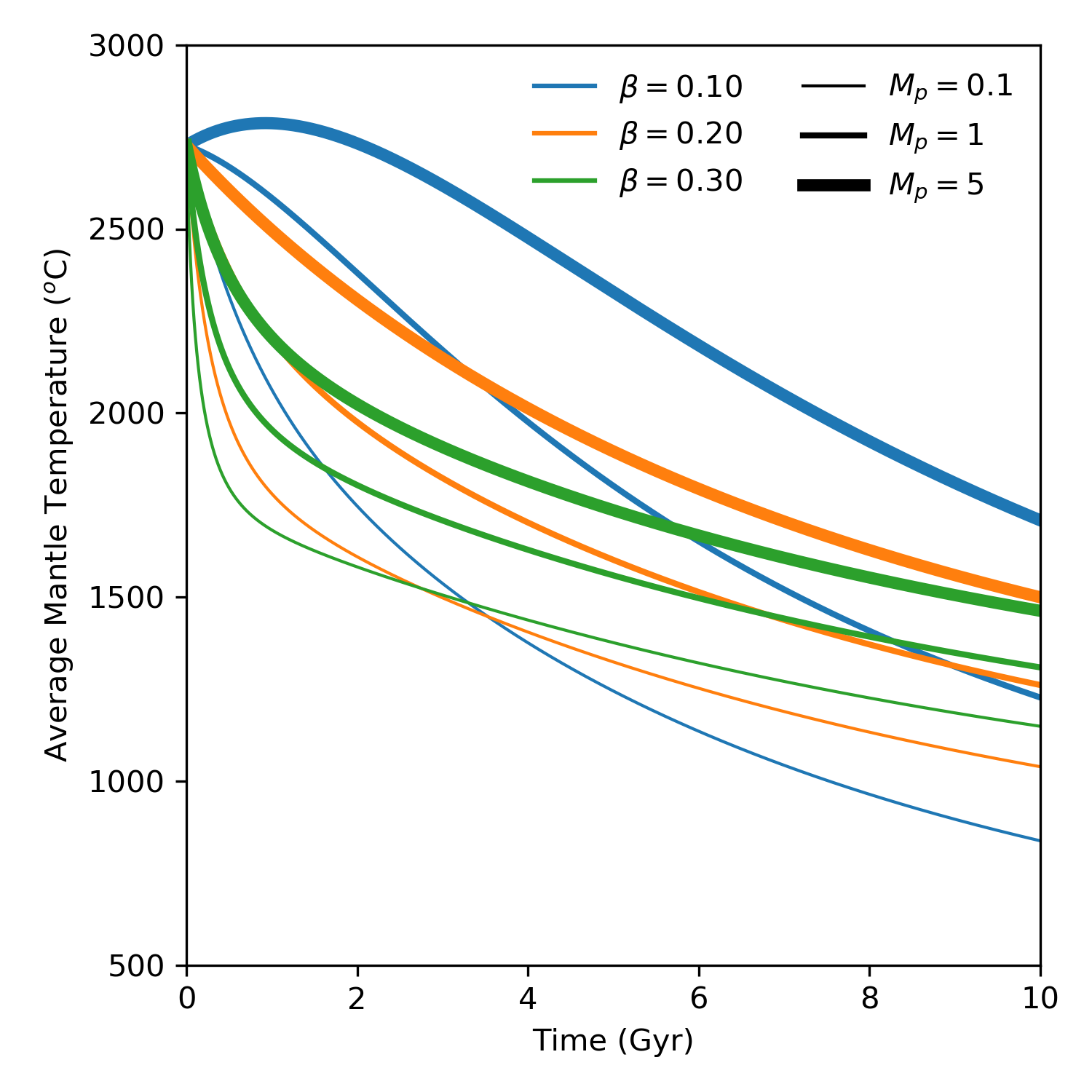}
      \caption{}
      \label{fig:Paths_SameTmi}
    \end{subfigure}%
    \begin{subfigure}[b]{0.35\textwidth}
      \includegraphics[width=\linewidth]{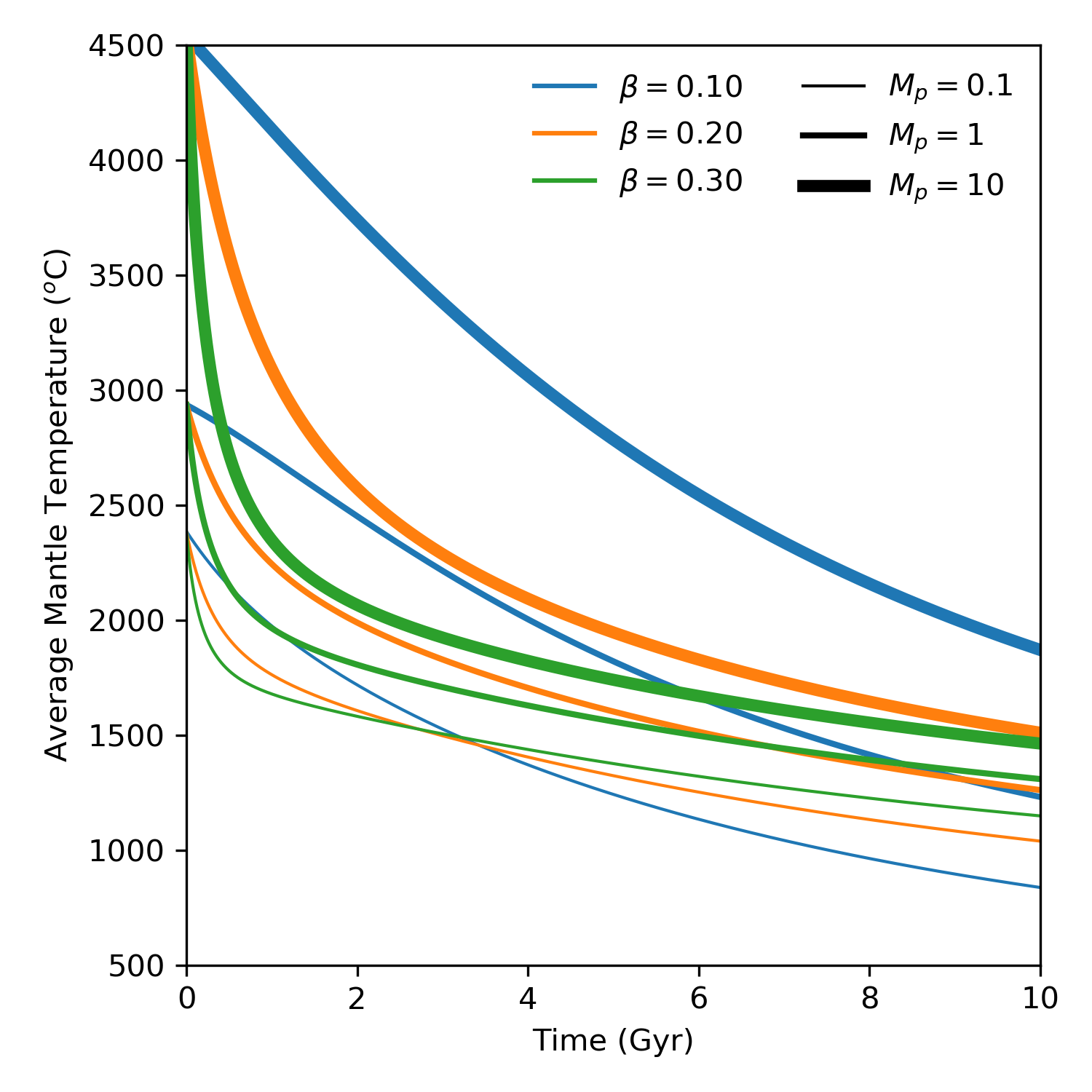}
      \caption{}
      \label{fig:Paths_DiffTmi}
    \end{subfigure}%
\caption{Sample thermal histories of average mantle temperature for CTMs (a and b) and STMs (c and d) that begin at the same (a and c) and different (b and d) temperatures.}
\label{fig:Paths}
\end{figure}

\section{Results}
Figure \ref{fig:Paths} shows sample thermal histories of different models and different starting temperatures. For low $\beta$ values, temperatures were considerably warmer for CTMs than STMs. This behavior was first noted by \citet{Mcnamara2001}. It occurs principally because CTMs have one initial value, mantle temperature, while STMs have effectively two boundary values, temperature and heat flux. This difference did not impact our principal conclusions. For a fixed tectonic cooling efficiency, small planet models cool faster than larger ones. Allowing for different plate tectonic cooling efficiencies produced more nuanced results. For example, a 5 $M_\oplus$ planet with $\beta=0.2$ had nearly the same temperature as an order of magnitude less massive planet after ten billion years of model evolution.  

\begin{figure}[h!]
  \centering
    \begin{subfigure}[b]{0.33\textwidth}
      \includegraphics[width=\linewidth]{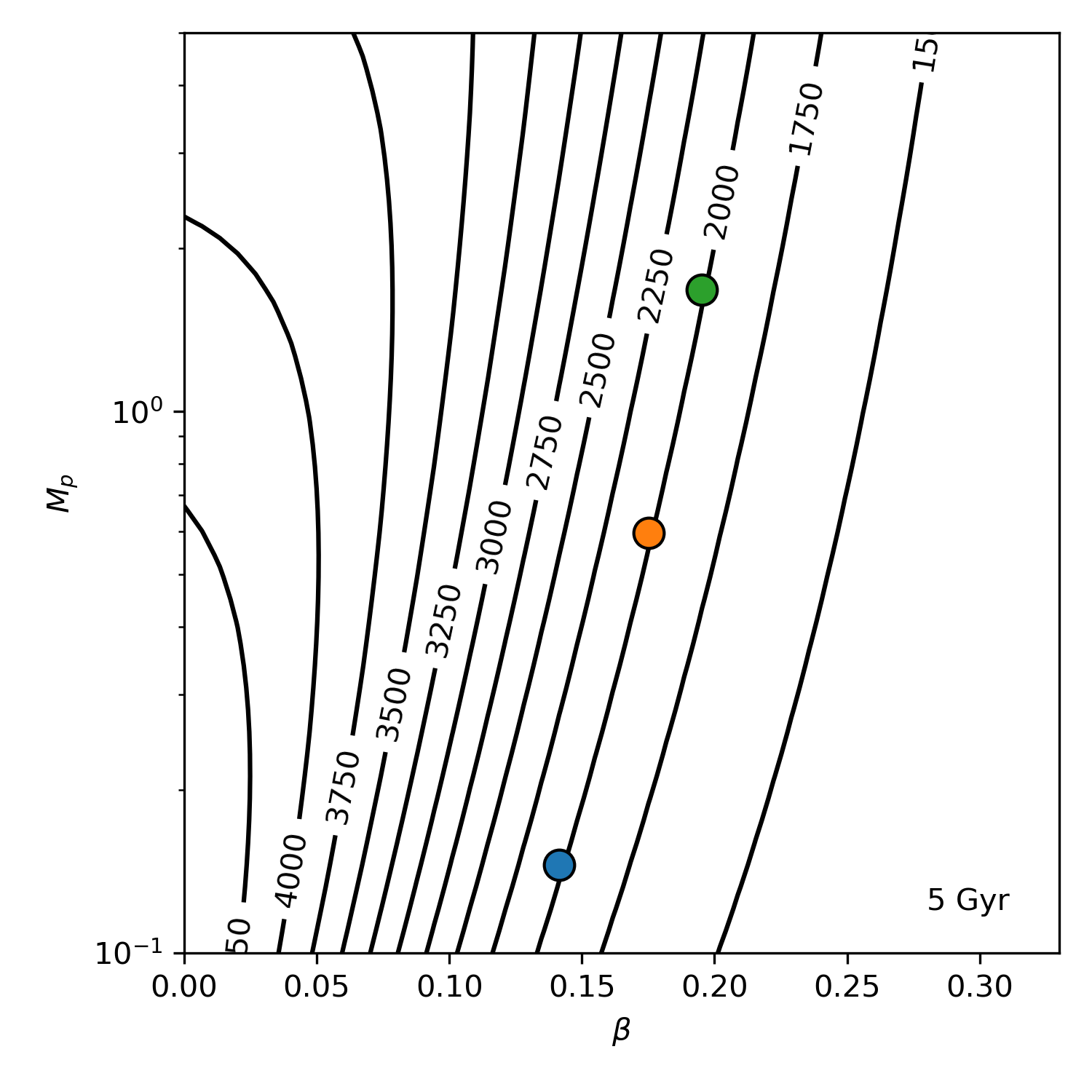}
      \caption{}
      \label{fig:Paths_SameTm_classic}
    \end{subfigure}%
    \begin{subfigure}[b]{0.33\textwidth}
      \includegraphics[width=\linewidth]{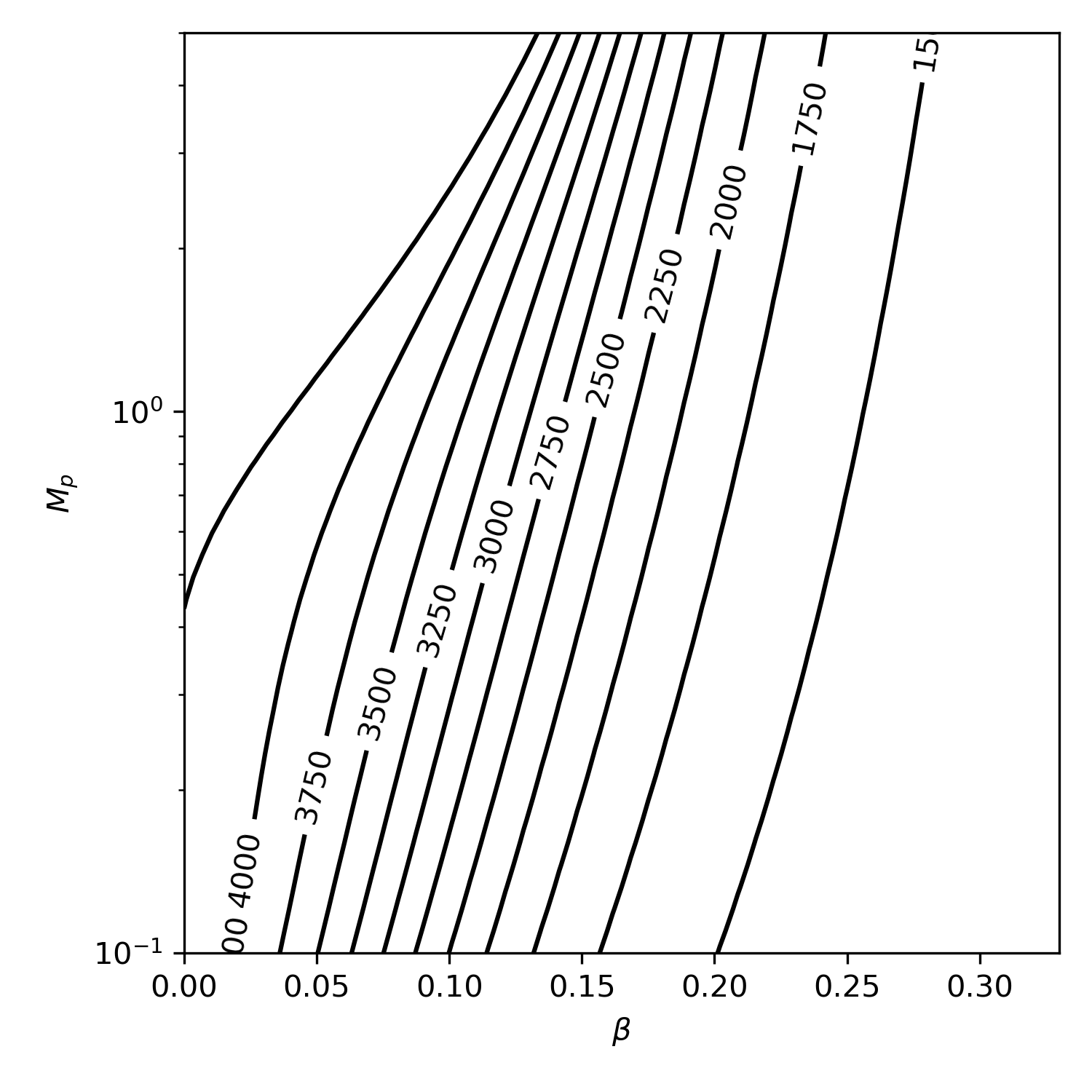}
      \caption{}
      \label{fig:Paths_DiffTm_classic}
    \end{subfigure}%
    \begin{subfigure}[b]{0.33\textwidth}
      \includegraphics[width=\linewidth]{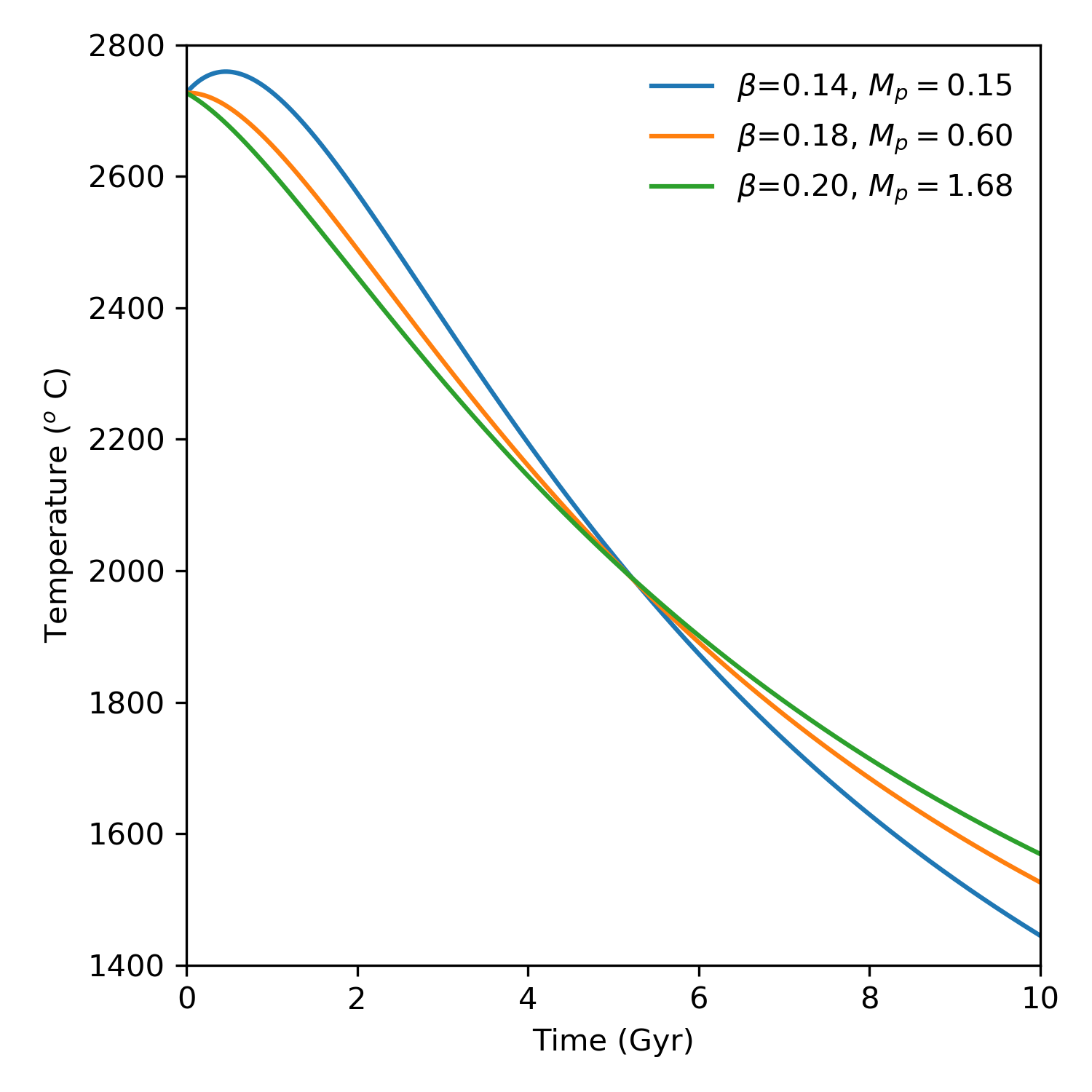}
      \caption{}
      \label{fig:Sample_Paths_CTM}
    \end{subfigure}%
    
    \medskip
    \begin{subfigure}[b]{0.33\textwidth}
      \includegraphics[width=\linewidth]{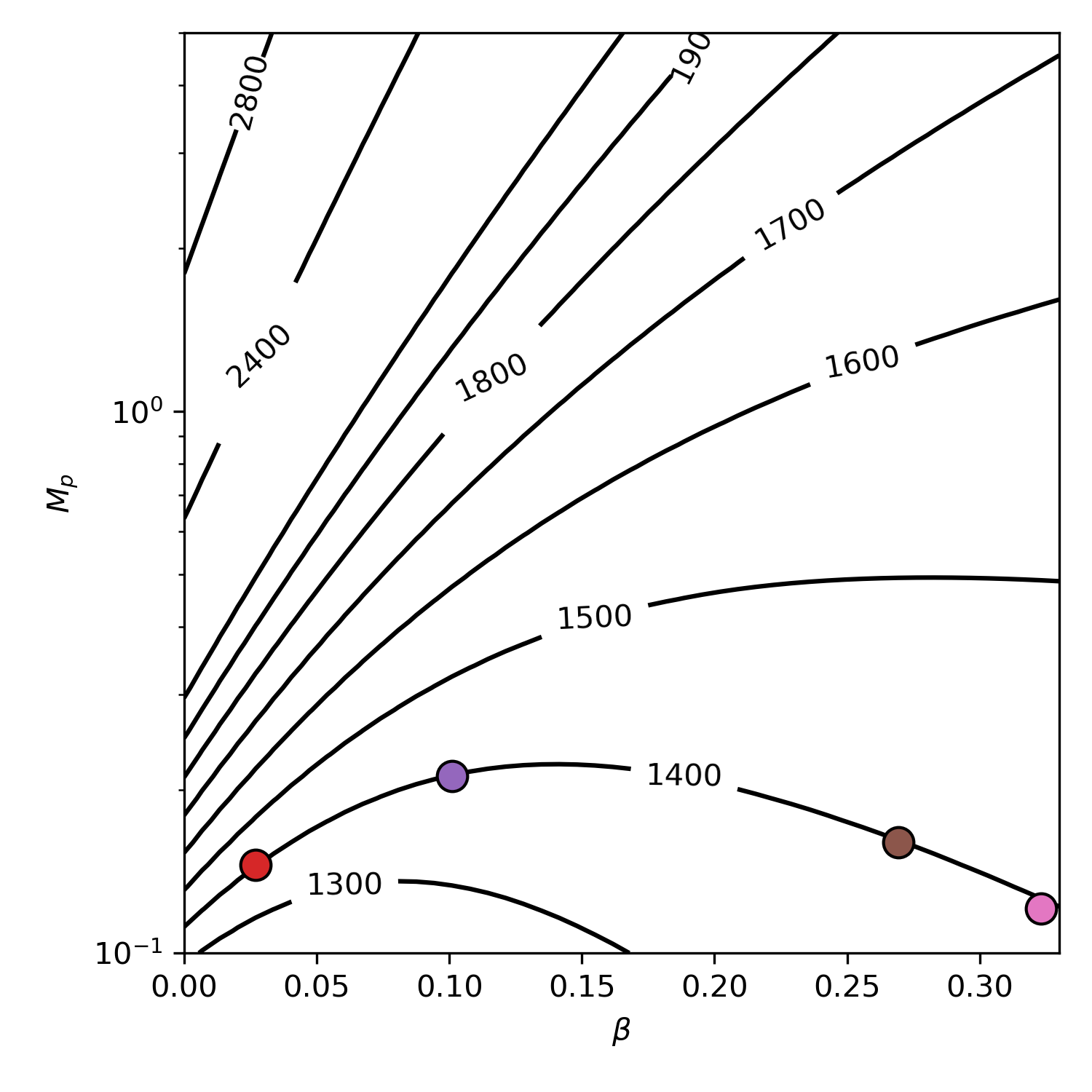}
      \caption{}
      \label{fig:Paths_SameTm}
    \end{subfigure}%
    \begin{subfigure}[b]{0.33\textwidth}
      \includegraphics[width=\linewidth]{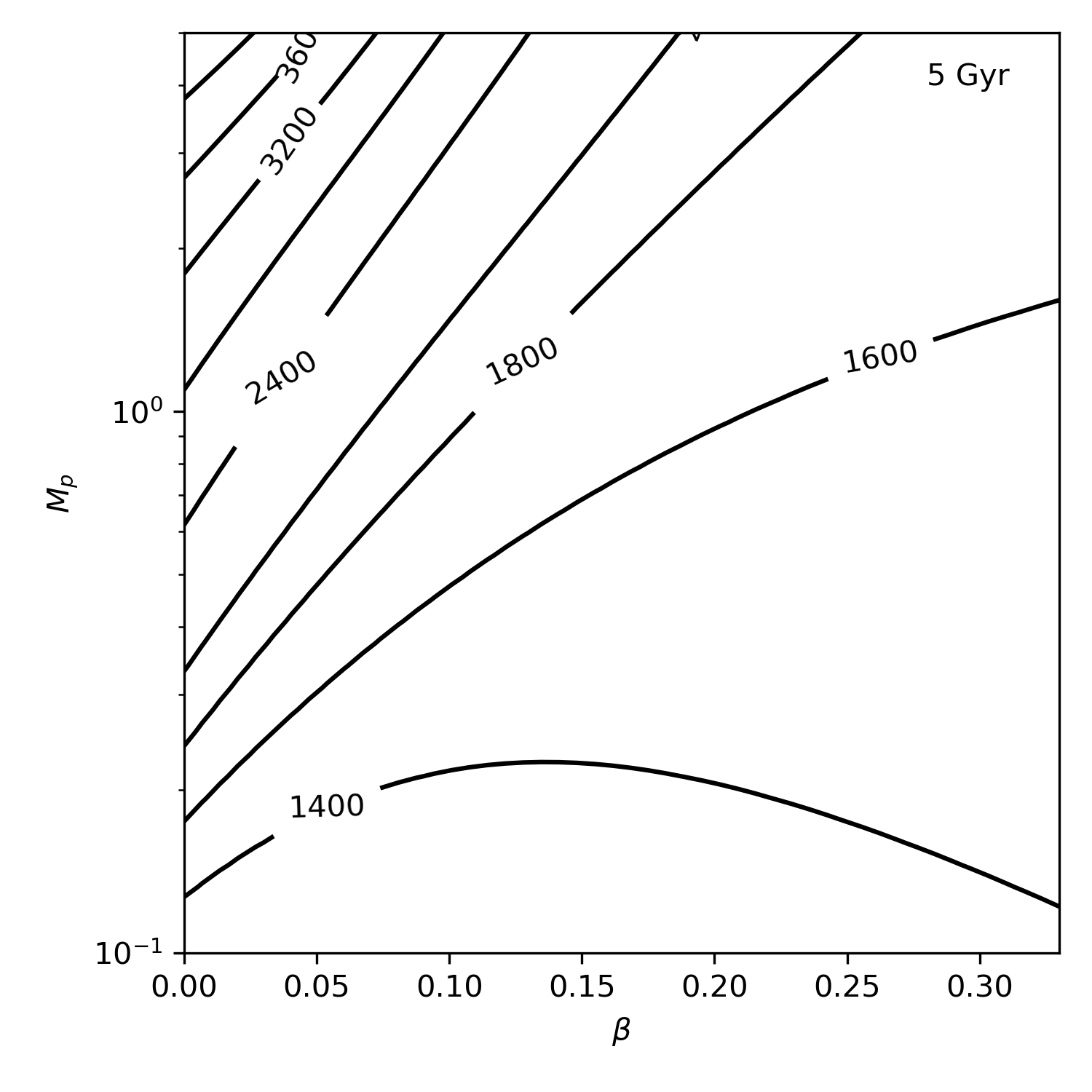}
      \caption{}
      \label{fig:Paths_DiffTm}
    \end{subfigure}%
    \begin{subfigure}[b]{0.33\textwidth}
      \includegraphics[width=\linewidth]{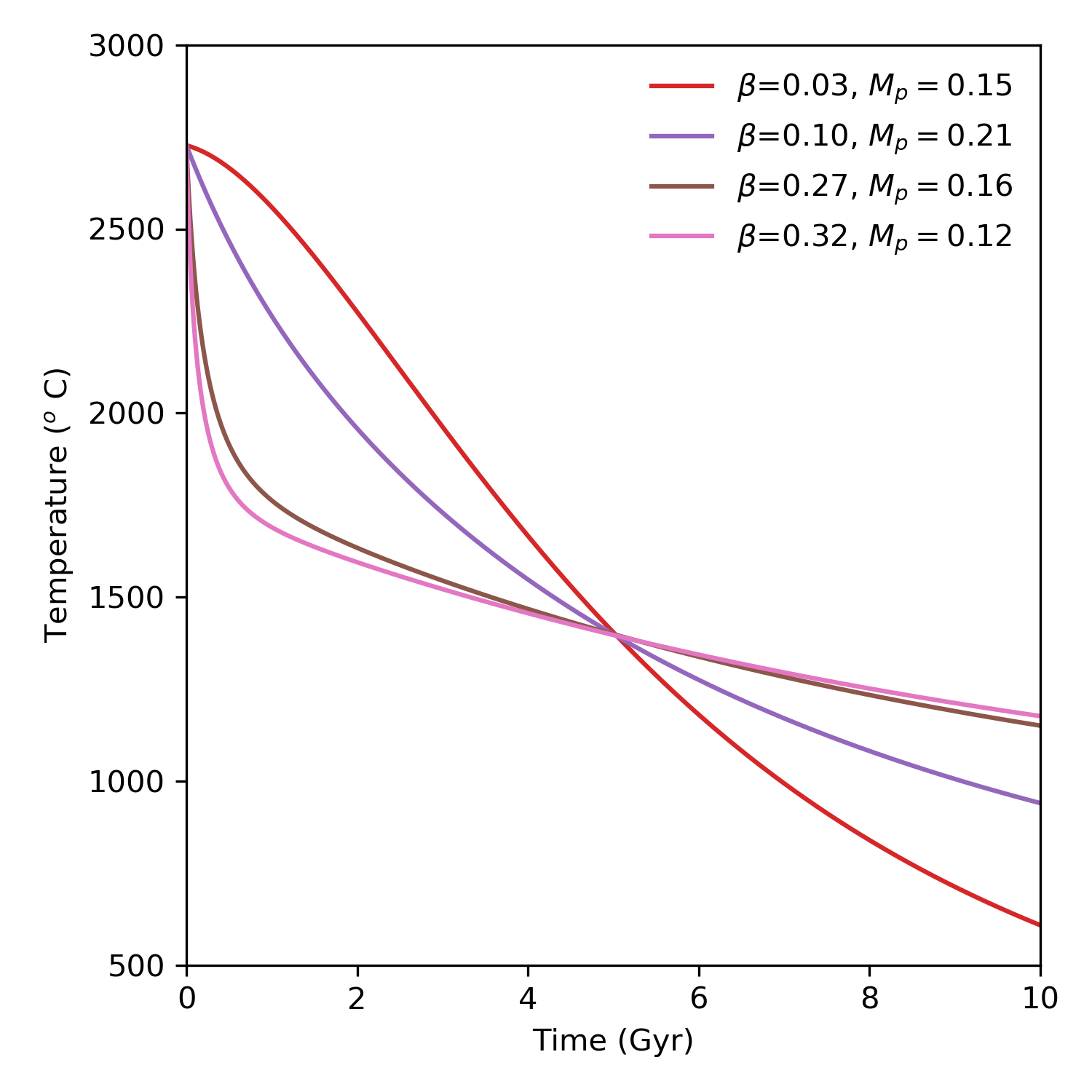}
      \caption{}
      \label{fig:Sample_Paths_STM}
    \end{subfigure}%
\caption{Contoured mantle temperatures at 5 Gyr for CTMs (a and b) and STMs (d and e) with the same (a and c) and different (b and d) initial mantle temperatures. Sample paths from this space demonstrating smaller planets can cool more slowly than larger ones (c and f).}
\label{fig:Contours_Tmi}
\end{figure}

Figure \ref{fig:Contours_Tmi} shows contoured mantle temperatures after five billion years of model time plotted in planetary mass and cooling efficiency space. Models on the same contour have cooled to the same temperature. The contours show that differences in plate tectonic cooling efficiencies allowed planets of different masses/sizes to be at the same temperature, i.e., small planets can cool to the same temperature as larger planets over time. Figures \ref{fig:Sample_Paths_CTM} and \ref{fig:Sample_Paths_STM} demonstrate this using sample thermal paths, which are color-coded to the parameter space shown in Figures \ref{fig:Paths_SameTm_classic} and \ref{fig:Paths_SameTm}. The CTM samples had similar cooling histories despite an order of magnitude difference in planetary mass. Similar behavior occurred for STM models with the addition that less massive (smaller) planets could remain significantly warmer than more massive ones for five billion years of model time. 

\section{Discussion and Conclusions}
The thermal history of a terrestrial planet affects its volcanic and geologic history. Volcanic/geologic history, in turn, affects the cycling of volatiles between a planet's interior and surface reservoirs, which is a critical factor in determining whether liquid water can exist at the surface of a planet over geological time \citep{Walker1981,Berner1983, Kasting1993,Kopparapu2014}. In addition to liquid water being key for life as we know it, life forms can also use a planets internal energy as a fuel source for their survival \citep{Baross1985a,Jannasch1985}. For these reasons, the solid body thermal evolution of a terrestrial planet has had a long standing connection to astrobiology. The discovery of terrestrial exoplanets has reinvigorated interest in that connection and in thermal history models. The discovery of terrestrial exoplanets larger and more massive than the Earth kick-started thinking about how differences in planetary size could affect a planets thermal history and, by association, life potential \citep{Valencia2007}. 

A first-wave of research into planetary size effects on geological history focused on whether larger planets would be more or less likely to have plate tectonics  \citep[e.g.,][]{Valencia2007,ONeill2007}. The focus was on the initiation of plate tectonics. That is, would internal energy overcome rock strength such that plate margins could be generated. Although an interesting problem, the cooling efficiency of plate tectonics does not depend solely one whether a planets internal energy sources can overcome rock strength to initiate plate subduction. It also depends on the source(s) of resistance to plate motions after plate tectonics is established. As discussed, that remains debated for the Earth and exoplanets with plate tectonics can have different cooling efficiencies. Allowing for this leads to trade-offs between a planets size/mass and tectonic cooling efficiency. A principal result is that planets smaller than the Earth, and of the same absolute age, can remain and geologically and volcanically active.

We have only considered differences in cooling efficiency for a particular tectonic mode (i.e., plate tectonics). Other tectonic regimes, such as episodic and stagnant lid, will further increase the possibility that planets of the same size as Earth may not have the same interior temperatures and/or that planets smaller than the Earth may have hotter interiors. Within our own solar system,  it has been argued that Venus may have liquid magma at the base of its mantle \citep{ORourke2020}. This suggests the interior of Venus may be hotter than Earth, despite the two planets having similar size and mass. Mars is considerably smaller and less massive than the Earth, yet estimates of its potential temperature are similar to Earth \citep{Filiberto2017}. In addition, \citet{Ruiz2011} argued that the Martian mantle experienced recent warming. An added effect that could allow small planets to remain geologically active is tidal locking. In some cases, tidal heating may be the dominant heat source in the mantle. Rocky bodies in that setting may maintain the same interior temperature for billions of years. With a large enough volatile inventory, this could provide steady, persistent outgassing of life essential elements \citep{Driscoll2015}. 

By looking at a range of plate tectonic cooling efficiencies, we have shown that smaller planets can cool slower than larger ones. This implies that, to the degree that geological activity is critical for planetary habitability, exoplanets smaller than the Earth, and of the same age or older, should not be down weighted in target selection strategies. An added implication is that planets sharing a range of Earth characteristics, including absolute age, can be at different times in their geological lifetimes -- the time window over which a planet can remain geologically active. To the degree that variations in volcanic/geologic/tectonic activity over time have influenced the evolution of life on Earth, this suggests that we should anticipate that Earth-like exoplanets, of the same age as Earth, need not be at the same evolutionary stages. 

\bibliography{SvLRefs}

\end{document}